\documentclass{article}
\usepackage{spconf,amsmath,graphicx, amssymb}
\usepackage{subcaption}
\usepackage{bm}
\usepackage{multirow}
\usepackage{booktabs}
\usepackage{tikz}
\usepackage{float}
\usepackage{hyperref}
\newcommand{\RNum}[1]{\uppercase\expandafter{\romannumeral #1\relax}}

\title{Personalized Speech Enhancement: New Models and\\ Comprehensive Evaluation}
\name{Sefik Emre Eskimez, Takuya Yoshioka, Huaming Wang, Xiaofei Wang, Zhuo Chen {\normalfont{and}} Xuedong Huang}
\address{Microsoft, One Microsoft Way, Redmond, WA, USA \\{\small \texttt{\{seeskime, tayoshio, huawang, xiaofewa, zhuc,  xdh\}@microsoft.com}}}

\begin{document}
\ninept
\maketitle

\begin{abstract} 
Personalized speech enhancement (PSE) models utilize additional cues, such as speaker embeddings like d-vectors, to remove background noise and interfering speech in real-time and thus improve the speech quality of online video conferencing systems for various acoustic scenarios. In this work, we propose two neural networks for PSE that achieve superior performance to the previously proposed VoiceFilter. In addition, we create test sets that capture a variety of scenarios that users can encounter during video conferencing. Furthermore, we propose a new metric to measure the target speaker over-suppression (TSOS) problem, which was not sufficiently investigated before despite its critical importance in deployment. Besides, we propose multi-task training with a speech recognition back-end. Our results show that the proposed models can yield better speech recognition accuracy, speech intelligibility, and perceptual quality than the baseline models, and the multi-task training can alleviate the TSOS issue in addition to improving the speech recognition accuracy.
\end{abstract}
\begin{keywords}
Speech enhancement, personalized speech enhancement, speaker embedding, automatic speech recognition, perceptual speech quality
\end{keywords}

\section{Introduction}
\label{sec:intro}
As a result of the COVID-19 pandemic, people and organizations are increasingly relying on digital technologies to stay connected and conduct business~\cite{pandey2020impact}. The usage of audio and video communication tools has surged since the pandemic started, and it is now more important than ever to provide high-quality speech audio even in acoustically adverse environments. 

Deep neural networks (DNNs) have shown promising results compared with the traditional signal processing methods for single-channel speech enhancement (SE)~\cite{choi2018phaseaware,phasen,hu2020dccrn,eskimez2021human}. The DNN models can remove the background noise in various environments and thus improve perceptual speech quality and intelligibility.

However, the DNN-based SE models let through all speakers present in the user's environment since there is no clue for the models to determine whom to attend. 
This limitation of the unconditional SE models, i.e.,  not filtering out the interfering speakers, can be undesirable and distracting in scenarios where the target user shares the space with other people or family members. Furthermore, it can potentially lead to serious privacy issues. Personal information of the household members may leak through the video/audio communication tools. In addition, confidential business information could also be unintentionally leaked for the same reason. 

Personalized SE (PSE) models can address these issues and have started gaining attention from researchers recently~\cite{giri2021personalized}. PSE combines the task of target speaker extraction~\cite{delcroix2018single,wang2019voicefilter,wang2020voicefilter} or speech separation~\cite{isik2016single,hershey2016deep,chen2017deep,chen2021conformer} and speech enhancement~\cite{choi2018phaseaware,phasen,hu2020dccrn,eskimez2021human}, and removes both the background noise and interfering speakers in real-time by using the cue obtained from the enrollment audio of the target speaker. 

The biggest practical challenge of PSE is target speaker over-suppression (TSOS). Due to the statistical nature of the speaker embedding, the PSE models, especially the causal ones, can sometimes get confused and falsely remove the target speaker's voice in addition to the background noise and interfering speakers. 
This phenomenon was reported in \cite{wang2020voicefilter}, which considered only automatic speech recognition (ASR) accuracy. However, the TSOS issue has a much more negative impact on human communication. It can be disruptive if the target speaker's voice is occasionally suppressed, even for a short period. A method to measure the TSOS reliably and a measure to counteract it are indispensable. 

In this paper, we propose two DNNs for PSE: 1) we convert the deep complex convolution recurrent network (DCCRN) model of \cite{hu2020dccrn}, which was proposed for unconditional SE, to PSE, and we name it personalized DCCRN (pDCCRN); 2) we also propose a new architecture that yields better performance than pDCCRN, namely personalized deep convolution attention U-Net (pDCATTUNET). Besides, we conduct comprehensive evaluation by using test sets encompassing different aspects of the PSE models. Furthermore, we propose a metric called \textit{TSOS measure} to quantify the degree of over-suppression and elucidate when the problem tends to happen. 
Finally, we adapt ASR-based multi-task (MT) training~\cite{eskimez2021human} to PNS and show that it can reduce the frequency of TSOS.


\section{Related Work}
\label{sec:relwork}

We define PSE as enhancing only the target speaker's voice while removing any other sounds present in the input signal, including interfering speakers and environmental noises. This definition is different than~\cite{personalized2021}, which defines PSE as fine-tuning an unconditional SE model on the target speaker's data without explicitly removing other speakers. The closest work in the literature is \cite{giri2021personalized}, which proposed Personalized PercepNet that runs in real-time and can remove the background speakers in addition to the noise. 

Another related field is target speaker extraction. Delcroix et al.~\cite{delcroix2018single} investigated the SpeakerBeam for single-channel target speaker extraction. SpeakerBeam utilizes the adaptation data of the target speaker for isolating the speaker by using an adaptation layer. They showed promising results for overlapped speech ASR. Wang et al. proposed VoiceFilter~\cite{wang2019voicefilter,wang2020voicefilter} to extract the target speaker for improving ASR accuracy. Their network is conditioned on a d-vector and includes convolution, LSTM, and linear layers. The network predicts the soft mask and applies it to the noisy magnitude spectrogram. Sato et al.~\cite{sato2021should} proposed a rule-based switching method to use target speaker extraction when overlapped speech is detected. They showed that this approach improved the ASR performance when there were interfering speakers. Another method was proposed by Xiao et al.~\cite{xiao2019single} which
leverages enrollment voice snippets of both the target speaker and interfering speakers. 

These previous studies focused only on either ASR accuracy or perceptual quality.
In addition, except for Personalized PercepNet, these works did not consider real-time processing.  
In contrast, we focus on real-time systems that improve both ASR accuracy and perceptual quality. 
We also extensively investigate the TSOS problem, which has yet to be sufficiently studied despite its importance for practical usage, and show that multi-task learning using ASR can significantly mitigate this problem. 


\section{Proposed Method}

Our previous experiments~\cite{eskimez2021human} showed that DCCRN~\cite{hu2020dccrn} achieved high-quality SE in terms of both ASR accuracy and perceptual quality. 
Therefore, we start our model architecture exploration by converting the unconditional DCCRN model to a personalized one that accepts a speaker d-vector as additional input. 
The d-vector extracted from an enrollment audio piece, or a speaker embedding, is supposed to represent the acoustic characteristics of the target speaker~\cite{zhou2021resnext}.
Then, we further propose a new model architecture based on self-attention~\cite{vaswani2017attention}. 
All the models described in this section are causal; they only process the current and past frames and, therefore, can actually be run in real-time on PCs. 

\subsection{Personalized DCCRN (pDCCRN)}
Conversion of the DCCRN requires the introduction of the d-vector as an input to the original architecture, and we aim for minimal change to preserve the unconditional model's performance. To achieve this, we explored a few configurations and obtained the best results when concatenating the d-vectors to the input of DCCRN's complex LSTM layer. Specifically, we concatenate the d-vectors to the real and imaginary parts of the tensors coming from the last layer of the encoder and feed the concatenated tensor to the complex LSTM. This modification increases the input size of the complex LSTM layer only; Therefore, the additional computational cost is minimal. We refer to this model as \textit{pDCCRN} for the rest of the paper.

\begin{figure}[t!]
  \centering
  \includegraphics[width=0.5\textwidth]{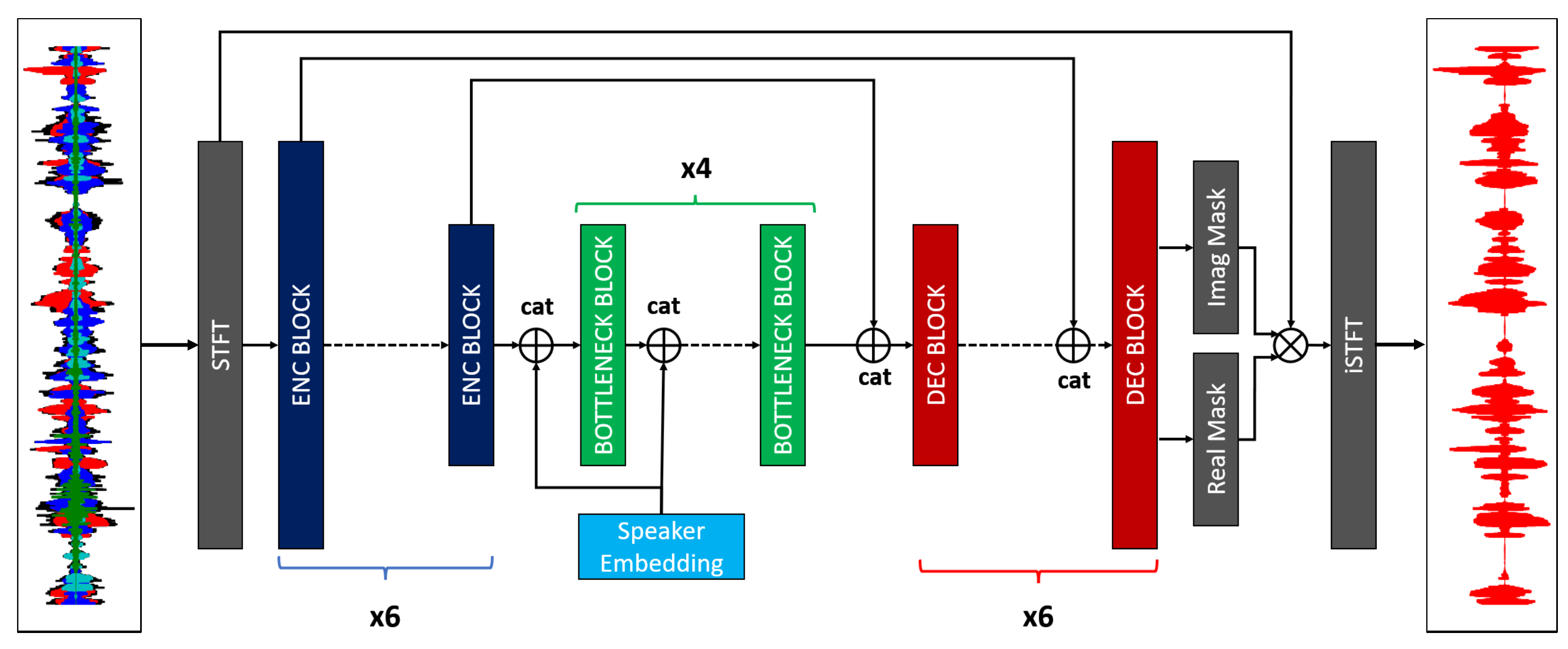}
  \vspace{-0.3cm}
  \caption{Proposed PSE model is shown. ``Cat'' stands for concatenation. $\bigotimes$ stands for complex multiplication. }
  \vspace{-0.5cm}
  \label{fig:cconvatt}
\end{figure}

\begin{figure}[t!]
  \centering
  \includegraphics[width=\columnwidth]{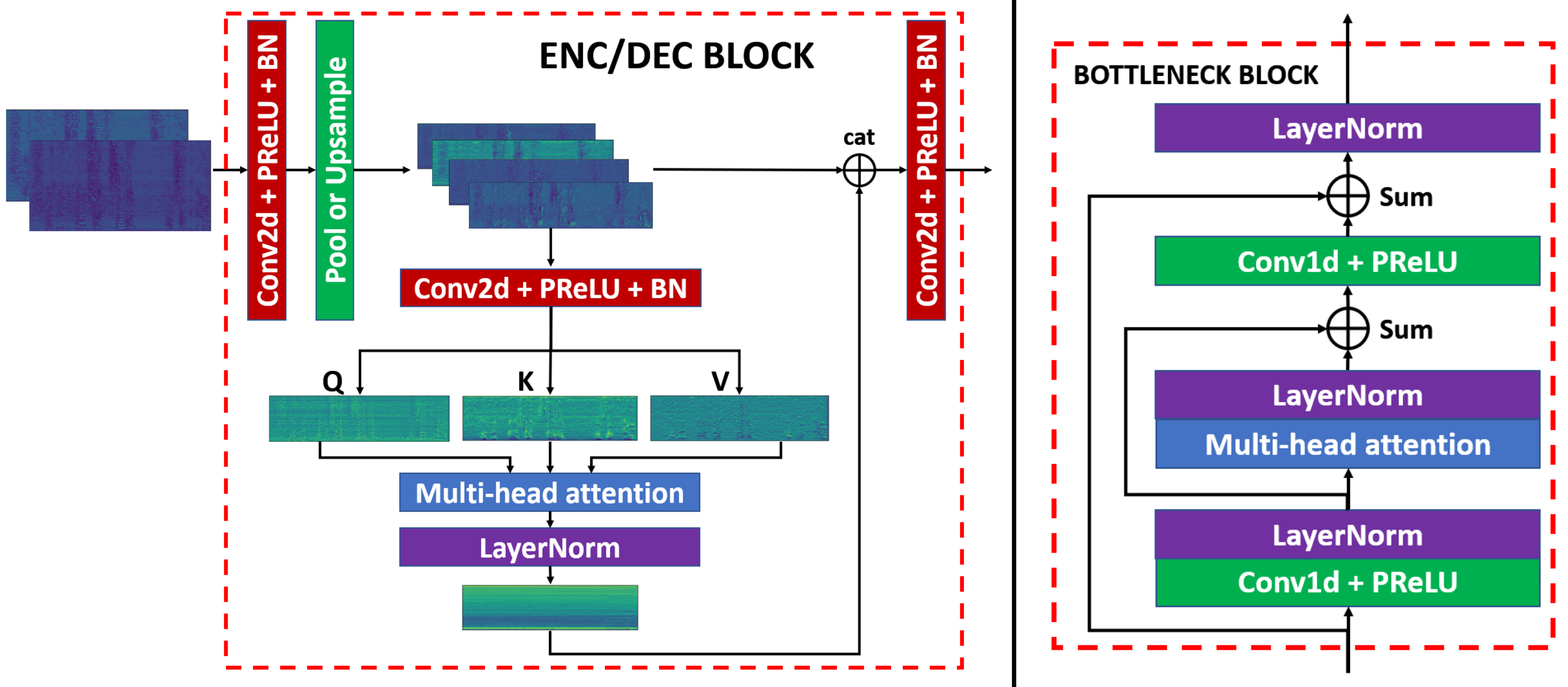}
  \vspace{-0.5cm}
  \caption{Diagram of pDCATTUNET blocks is shown. ``Q'', ``K'', ``V'' and ``BN'' stands for query, key, value, and batch normalization, respectively. }
  \vspace{-0.5cm}
  \label{fig:modules}
\end{figure}

\subsection{Personalized Deep Convolution Attention U-Net  (pDCATTUNET)}



DCCRN relies on the bottleneck layer (LSTM) to model the temporal dependencies. However, we can incorporate temporal modeling in the convolutional layers by leveraging a multi-head attention module. For example, Ashutosh et al.~\cite{pandey2021dense} proposed a dense convolutional network with self-attention. In our network, we infuse multi-head attention in convolutional blocks, which differ from~\cite{pandey2021dense} in terms of implementation.  The model design is illustrated in Figure~\ref{fig:cconvatt}. 

\textbf{Encoder/Decoder Blocks: } Our encoder and decoder architectures include convolution layers and are supported by multi-head attention~\cite{vaswani2017attention}. The encoder/decoder blocks feed their input to a convolution layer with a 2-D kernel, and it is followed by a parametric rectified linear unit (PReLU) and batch normalization (BN). For simplicity, we name the combination of convolution layer followed by PReLU and BN as \textit{convolution block} for the rest of the paper. These layers are followed by a max-pooling layer for encoder blocks and a nearest-neighbor upsampling layer for decoder blocks. The downsampling/upsampling operations are applied to the frequency dimension. Another convolution block processes these intermediate results and outputs three channels. These intermediate embeddings are used as the query, key, and value for $MultiHead(Q, K, V)$~\cite{vaswani2017attention} with a causal attention mask. The output of the multi-head attention module is normalized by a normalization layer and concatenated to the intermediate embeddings coming from the first convolution block. A third convolution block takes the concatenated embeddings as input and sends its output to the next layer. The details are shown in Figure~\ref{fig:modules}. 

\textbf{Bottleneck Blocks: } The success of the transformer and its variants for speech processing~\cite{wang2020transformer,gulati2020conformer,chen2021conformer} inspired the design of the bottleneck layers. Instead of an RNN-based bottleneck, we design blocks using convolution layers with 1-D kernels and multi-head attention modules. Input to the bottleneck block is processed by a convolution layer followed by PReLU and layer normalization. Then, the intermediate results are fed to a multi-head attention module, followed by layer normalization. There is a skip connection (summation) between the input and output of the multi-head attention + layer normalization block. Skip connection is followed by another convolution layer + PReLU layer. It is followed by a second skip connection: the input of the bottleneck block is summed with the output of the second convolution block. Finally, another layer normalization layer is applied, and the results are sent to the next layer. 

We apply 1-D batch normalization to real and imaginary parts of the spectrogram after the STFT. The network takes the concatenation of the real and imaginary parts of the STFT as input feature, in channel dimension, resulting in a 4-D tensor including batch dimension. After predicting the real and imaginary parts of the mask, we apply the estimated complex ratio mask~\cite{williamson2015complex} to the original real and imaginary parts of the noisy spectrogram.

\subsection{Loss Function}
\cite{eskimez2021human} showed the effectiveness of a power-law compressed phase-aware (PLCPA) loss function for both ASR accuracy and perceptual quality. 
It is defined as follows:
\begin{equation}\label{loss}
\begin{split}
\mathcal{L}_{a}(t, f) &=  
\left||S(t, f)|^p-|\hat{S}(t, f)|^p\right|^2 \\
\mathcal{L}_{p}(t, f) &= \left||S(t, f)|^pe^{j\varphi (S(t, f))}-|\hat{S}(t, f)|^pe^{j\varphi (\hat{S}(t, f))}\right|^2 \\
\mathcal{L}_{PLCPA} &= \frac{1}{T} \frac{1}{F}\sum_{t}^{T} \sum_{f}^{F} (\alpha~\mathcal{L}_{a}(t, f) + (1-\alpha)~\mathcal{L}_{p}(t, f)),
\end{split}
\end{equation}
where $S$ and $\hat{S}$ are the estimated and reference (i.e., clean) spectrograms, respectively. $t$ and $f$ stand for time and frequency index, while $T$ and $F$ stand for the total time and frequency frames, respectively. Hyper-parameter $p$ is a spectral compression factor and is set to 0.3. Operator $\varphi$ calculates the argument of a complex number. $\alpha$ is the weighting coefficient between the amplitude and phase-aware components. 

To alleviate the over-suppression (OS) issue, we adapt the asymmetric loss described in~\cite{wang2020voicefilter} to the amplitude part of the loss function defined by~\eqref{loss}. The asymmetric loss penalizes the T-F bins where the target speaker's voice is removed. The power-law compressed phase-aware asymmetric (PLCPA-ASYM) loss function is defined as follows:
\begin{equation}\label{loss_asym}
\begin{split}
h(x) &=\begin{cases}
    0, & \text{if $x\le0$},\\
    x, & \text{if $x>0$},
  \end{cases} \\
\mathcal{L_{OS}}(t, f) &=  
\left|h(|S(t, f)|^p-|\hat{S}(t, f)|^p)\right|^2 \\
\mathcal{L}_{PLCPA-ASYM} &= \mathcal{L}_{PLCPA} + \beta \frac{1}{T} \frac{1}{F}\sum_{t}^{T} \sum_{f}^{F} \mathcal{L_{OS}}(t, f),
\end{split}
\end{equation}
where $\beta$ is the positive weighting coefficient for $L_{OS}$. 

Furthermore, we perform multi-task (MT) training with ASR using a frozen back-end model as the sub-task   
to update the parameters of the PSE network. 
This was inspired by \cite{eskimez2021human}, which applied the MT training to the unconditional SE and showed superior ASR performance.
In our experiments, 
since we did not have the speaker enrollment audio for the ASR training data, we extracted the d-vectors directly from the noisy training utterances. The MT loss is denoted as $\mathcal{L}_{MT}$. See \cite{eskimez2021human}
for details. 
\section{Experiments}

This section reports the results of the experiments that we conducted to analyze various aspects of the proposed PSE models. For the d-vector extraction, we used the pre-trained Res2Net speaker ID model described in~\cite{zhou2021resnext}.

\subsection{Training and Validation Data} \label{sec:trainvaldata}
We used the clean speech data of the deep noise suppression (DNS) challange~\cite{reddy2021interspeech}. The dataset consists of around 544 hours of speech samples from the LibriVox corpus~\cite{kearns2014librivox} that were obtained by selecting high mean opinion score (MOS) samples. We employed the noise files included in the Audioset~\cite{gemmeke2017audio} and Freesound~\cite{fonseca2017freesound} datasets and room impulse response (RIR) files simulated by the image method to create noisy mixtures. We split the noise and RIR files into training, validation, and test sets. A 60\% portion of the dataset contained samples including the target speaker, interfering speaker, and noise; the other 40\% contained samples comprising the target speaker and noise. The latter subset was included to maintain good performance for scenarios with no interfering speakers (i.e., the most typical SE scenario), which is crucial in practice. We assumed that the target speaker would be closer to the microphone than the interfering speaker. Therefore, we placed the target speaker randomly between 0 to 1.3 meters away from the microphone while placing the interfering speaker more than 2 meters away. We simulated 2000 and 50 hours of data for the training and validation, respectively. 

\subsection{Test Sets}
We created our test sets to measure the PSE models' performance for three practically relevant scenarios: 
1) the scenario of primary interest (target + interfering speakers + noise), 2) the scenario where there are no interfering speakers (target speaker + noise), and 3) the scenario where there are neither interfering speakers nor noise (target speaker only). For brevity, we name scenarios 1), 2), and 3) as ``TS1'', ``TS2'', and ``TS3'', respectively. TS2 and TS3 are benchmark test sets with which we can compare the PSE models with the unconditional SE models in a fair way. All the test sets contained reverberation. 

We used two clean speech sources with different acoustic properties for creating the test sets: 1) internal conversational recordings (ICR) and 2) the voice cloning toolkit (VCTK) corpus. 
The ICR data contained meeting audio recordings of 13 speakers obtained with headset microphones. The long-duration audio is segmented at silence positions based on manual transcription. We only kept the segments that included a single speaker. The average duration of the recordings was 9 seconds, and the total duration was  and RIRs mentioned in Section~\ref{sec:trainvaldata}. For each speaker, we used around 60 seconds of held-out enrollment audio to extract the d-vectors.

The VCTK corpus includes 109 speakers with different English accents. For each speaker, we set aside 30 samples and used them to extract the speaker's d-vector. Then, we simulated the noisy mixtures using the test noise and RIRs with the rest of the files. These noisy mixtures were concatenated to generate a single long audio file for each speaker. The average duration of the files was 27.5 minutes. This test set served the purpose of evaluating the models with long-duration audio to which the models were not exposed during training and under changing environmental conditions. 

\begin{table*}[ht!]
  
  \caption{Experimental results for ICR and VCTK datasets using three testing conditions. TS1 includes target, interfering speaker, and noise, TS2 includes target speaker and noise, and TS3 includes only the target speaker. All conditions include reverberation. For WER, DEL, and TSOS metrics, the lower is better. For DNSMOS and STOI metrics, the higher is better. The best value for each column is marked with a bold font, excluding ``No Enhancement'' row. Models are trained with $\mathcal{L}_{PLCPA}$ unless marked otherwise.}
  \centering
 \resizebox{\textwidth}{!}{
\begin{tabular}{llcccccccccccccc}
\hline
\multicolumn{1}{c}{\multirow{3}{*}{Model}} &  & \multicolumn{14}{c}{\textbf{ICR} - Short-Duration Conversational Audio}                                                                                                                                                                                                                                                                                                                              \\ \cline{3-16} 
\multicolumn{1}{c}{}                       &  & \multicolumn{5}{c}{TS1}                                                                                                                   &  & \multicolumn{5}{c}{TS2}                                                                                                                   &  & \multicolumn{2}{c}{TS3}                                 \\ \cline{3-7} \cline{9-13} \cline{15-16} 
\multicolumn{1}{c}{}                       &  & \multicolumn{1}{c}{WER} & \multicolumn{1}{c}{DEL} & \multicolumn{1}{c}{DNSMOS} & \multicolumn{1}{c}{STOI (\%)} & \multicolumn{1}{c}{TSOS (\%)} &  & \multicolumn{1}{c}{WER} & \multicolumn{1}{c}{DEL} & \multicolumn{1}{c}{DNSMOS} & \multicolumn{1}{c}{STOI (\%)} & \multicolumn{1}{c}{TSOS (\%)} &  & \multicolumn{1}{c}{DEL} & \multicolumn{1}{c}{TSOS (\%)} \\ \cline{1-13} \cline{15-16} 
No Enhancement              && 36.36 & 9.55 & 3.03 & 71.33 & 0.00 && 20.82 & 6.50 &	3.05 &	74.96 &	0.00 && 3.21 & 0.00 \\ \hline
DCCRN                       && 40.12 & 11.38 & 3.64 & 79.85 & 9.01 && 25.43 & 7.81 & 3.77 & 85.70 & 7.32 && 3.27 & 1.47 \\ \hline
VoiceFilter                 && 31.55 & 10.78 & 3.42 & 81.89 & 11.91 && 26.36 & 8.81 & 3.60 & 84.54 & 11.45 && 3.44 & 2.89 \\ \hline
pDCCRN                      && 29.60 & 10.52 & 3.71 & 82.68 & 14.51 && 25.06 & 8.22 & 3.84 & 85.64 & 13.55 && 3.48 & 4.54 \\ 
\hspace{0.4cm} - $\mathcal{L}_{PLCPA-ASYM}$  && 28.34 & 9.05 & 3.71 & 84.45 & 6.80 && 24.20 & 7.50 & 3.84 & 86.93 & 6.05 && 3.42 & 1.62 \\
\hspace{0.4cm} - $\mathcal{L}_{MT}$  && 27.41 & 9.03 & 3.68 & 83.80 & 5.46 && 22.56 & 7.25 & 3.80 & 86.49 & 4.84 && 3.19 & 2.86 \\
\hspace{0.4cm} - $\mathcal{L}_{PLCPA-ASYM}$ + $\mathcal{L}_{MT}$  && 26.91 & 8.68 & 3.67 & 84.60 & 3.13 && 22.49 & 7.17 & 3.79 & 87.09 & 2.48 && 3.15 & 1.64 \\ \hline
pDCATTUNET                  && 25.78 & 8.04 & \textbf{3.78} & 85.58 & 9.53 && 22.67 & 6.69 & \textbf{3.90} & 87.55 & 9.02 && 3.35 & 2.56 \\  
\hspace{0.4cm} - $\mathcal{L}_{PLCPA-ASYM}$  && 25.43 & 7.65 & \textbf{3.78} & \textbf{86.10} & 6.93 && 22.47 & 6.42 & \textbf{3.90} & \textbf{88.10} & 6.20 && 3.21 & 1.65 \\ \hspace{0.4cm} - $\mathcal{L}_{MT}$  && \textbf{24.94} & 7.98 & 3.72 & 85.86 & 3.22 && 21.38 & 6.55 & 3.83 & 87.88 & 2.91 && 3.18 & 1.48 \\
\hspace{0.4cm} - $\mathcal{L}_{PLCPA-ASYM}$ + $\mathcal{L}_{MT}$  && 25.01 & \textbf{7.33} & 3.72 & 85.73 & \textbf{1.20} && \textbf{21.05} & \textbf{6.23} & 3.82 & 87.82 & 0.90 && \textbf{3.15} & \textbf{0.69} \\
\hline \hline
                                           &  & \multicolumn{14}{c}{\textbf{VCTK} - Long-Duration Audio}                                                                                                                                                                                                                                                                                                                             \\ \cline{3-16} 
\multicolumn{1}{c}{}                       &  & \multicolumn{5}{c}{TS1}                                                                                                                   &  & \multicolumn{5}{c}{TS2}                                                                                                                   &  & \multicolumn{2}{c}{TS3}                                 \\ \cline{3-7} \cline{9-13} \cline{15-16} 
\multicolumn{1}{c}{}                       &  & \multicolumn{1}{c}{WER} & \multicolumn{1}{c}{DEL} & \multicolumn{1}{c}{DNSMOS} & \multicolumn{1}{c}{STOI (\%)} & \multicolumn{1}{c}{TSOS (\%)} &  & \multicolumn{1}{c}{WER} & \multicolumn{1}{c}{DEL} & \multicolumn{1}{c}{DNSMOS} & \multicolumn{1}{c}{STOI (\%)} & \multicolumn{1}{c}{TSOS (\%)} &  & \multicolumn{1}{c}{DEL} & \multicolumn{1}{c}{TSOS (\%)} \\ \cline{1-13} \cline{15-16}
No Enhancement              && 43.03 & 3.95 & 2.92 & 78.87 & 0.00 && 13.35 & 1.97 &	2.98 & 85.00 & 0.00 && 0.77 & 0.00 \\ \hline
DCCRN                       && 54.64 & 4.15 & \textbf{3.50} & 85.41 & 5.13 && 19.11 & 2.81 & 3.63 & 92.31 & 5.33 && 0.81 & 2.42 \\ \hline
VoiceFilter                 && 37.94 & 6.24 & 3.15 & 85.41 & 8.37 && 18.75 & 3.54 & 3.38 & 91.18 & 5.31 && 0.96 & 0.53 \\ \hline
pDCCRN                      && 35.75 & 10.97 & 3.37 & 83.30 & 16.89 && 19.65 & 5.73 & 3.63 & 90.30 & 9.80 && 4.31 & 4.86 \\ 
\hspace{0.4cm} - $\mathcal{L}_{PLCPA-ASYM}$  && 35.48 & 7.43 & 3.41 & 86.18 & 7.73 && 19.04 & 4.72 & 3.62 & 91.19 & 5.94 && 1.85 & 1.61 \\
\hspace{0.4cm} - $\mathcal{L}_{MT}$  &&  34.06 & 4.58 & 3.40 & 87.35 & 3.60  && 15.72 & 2.60 & 3.57 & 92.04 & 3.22 && 0.82 & 0.57 \\
\hspace{0.4cm} - $\mathcal{L}_{PLCPA-ASYM}$ + $\mathcal{L}_{MT}$  && 32.90 & 5.22 & 3.36 & 87.70 & 2.93 && 15.92 & 2.95 & 3.52 & 92.21 & 2.69 && 0.94 & 1.12 \\ \hline
pDCATTUNET                  &&  31.72 & 5.50 & 3.46 & 87.93 & 7.47  &&  17.62 & 3.98 & \textbf{3.69} & 91.85 & 7.48  && 3.73 & 5.86 \\  
\hspace{0.4cm} - $\mathcal{L}_{PLCPA-ASYM}$  &&  \textbf{31.57} & 5.55 & 3.44 & 88.31 & 5.32  && 17.91 & 4.25 & 3.65 & 92.02 & 5.37  && 2.59 & 2.62 \\ 
\hspace{0.4cm} - $\mathcal{L}_{MT}$  &&  35.23 & \textbf{3.24} & 3.44 & \textbf{89.45} & 1.53  &&  \textbf{14.78} & \textbf{2.13} & 3.60 & \textbf{93.21} & 1.64  && 0.83 & 0.93 \\
\hspace{0.4cm} - $\mathcal{L}_{PLCPA-ASYM}$ + $\mathcal{L}_{MT}$  &&  33.85 & 3.31 & 3.45 & 89.43 & \textbf{1.07}  &&  14.96 & 2.22 & 3.60 & 93.18 & \textbf{0.88} && \textbf{0.81} & \textbf{0.30} \\ \hline
\end{tabular}
  }
  \label{tab:results}
\vspace{-0.5cm}
\end{table*}

\subsection{Evaluation Metrics and TSOS Measure}
We want to measure the ASR accuracy and perceptual quality improvements for the PSE models. Therefore, we rely on the word error rate (WER), deletion error rate (DEL), short-time objective intelligibility (STOI)~\cite{taal2011algorithm} and DNSMOS~\cite{gamper2019intrusive}. DNSMOS is a neural network-based mean opinion score (MOS) estimator that was shown to correlate well with the quality ratings by humans. 

We employ DEL in addition to WER since we want to measure the over-suppression (OS) of the target speaker. 
While it is correlated with the degree of OS, 
DEL alone was found to fail in capturing the OS of the target speaker perfectly. 
To complement DEL, we propose a new metric to measure the target speaker OS at the signal level. The metric is defined for each time frame as follows:  
\begin{equation}\label{tsos}
\begin{split}
\mathcal{TSOS}(t) &= \begin{cases}
    1, & \text{if $\sum_{f} \mathcal{L_{OS}}(t, f) > \gamma \sum_{f} |S(t, f)|^p$},\\
    0, & \text{otherwise},
  \end{cases} \\
\end{split}
\end{equation}
where $\gamma$ is a threshold value, $\mathcal{L_{OS}}$ represents the OS rate and defined in~\eqref{loss_asym}. 
With the frame-level $\mathcal{TSOS}$ measure, we can calculate the percentage of the OS frames, total OS duration, and maximum OS duration. Due to space limitations, we only report the mean TSOS percentage.


\subsection{Baseline Method and Implementation Details}
We compare our models with VoiceFilter~\cite{wang2019voicefilter}. We used the same parameters as described in~\cite{wang2019voicefilter}, except that we converted the VoiceFilter model to a causal one by making the paddings for convolution layers causal and the LSTM layers unidirectional. For DCCRN and pDCCRN, the numbers of convolution filters were $[16, 32, 64, 128 ,128, 128]$ for the encoder and decoder (in the reversed order) layers, and we set the kernel size to $5 \times 2$ and stride to $2 \times 1$ for each layer. The LSTM layer's hidden size was set to 128. For pDCATTUNET, the numbers of filters were $[32, 64, 128 ,128, 128, 128]$ for the encoder and $[128, 128, 128 ,64, \\ 32, 16]$ for the decoder. The hidden size of the bottleneck layer was set to 128. 
We selected the best checkpoint according to the validation loss. We set $\alpha$ in~\eqref{loss} and~\eqref{loss_asym} to 0.5 and 0.9, respectively. We set $\beta$ in~\eqref{loss_asym} to 1.0. Furthermore, we set $\gamma$ in~\eqref{tsos} to 0.1. 

\subsection{Results}
Table~\ref{tab:results} shows the experimental results for both ICR and VCTK datasets. 
Overall, pDCATTUNET yielded significantly better results for both datasets than pDCCRN and the baseline models in all performance metrics. For TS2, 
while pDCCRN and DCCRN yielded similar results in both ICR and VCTK, pDCATTUNET outperformed them, showing the effectiveness of this architecture even for the no-interfering-speaker case.  

The personalized models showed higher TSOS rates than the unconditional DCCRN model for all scenarios when they were trained with the same loss function. While the DEL rate of DCCRN is similar for all conditions of the ICR dataset, it is lower for the long-duration audio (VCTK) compared to the personalized models. Considering the TSOS results as well, we can conclude that for the long-duration audio, the personalized models suffer more OS. This might be due to VCTK containing samples with changing acoustic environments every few seconds or using 10 seconds samples during training. Using the asymmetric loss function significantly improved the TSOS rate and speech intelligibility for all cases. Besides, the MT training with $\mathcal{L}_{PLCPA}$ loss provided a more substantial improvement to the TSOS rate than the asymmetric loss. On the other hand, the MT training slightly reduced the perceptual quality, which is in agreement with our finding in~\cite{eskimez2021human}. Finally, the MT training and asymmetric loss combination achieved the best TSOS rate for both pDCCRN and pDCATTUNET.\footnote{Samples available at \scriptsize \url{https://eeskimez.github.io/pse-samples/}} 

It must be noted that for long-duration audio, the MT training seems to degrade the overlapped speech WER for the pDCATTUNET, although the DEL was reduced. Our analysis found that the insertion error was increased, which means that the suppression of the interfering speech was compromised. We suspect the reason is related to the usage of a single speaker for the ASR data during MT training. Adding interfering speech to ASR data might alleviate this issue.

\section{Conclusions}
In this work, we studied the personalized speech enhancement (PSE) problem, proposed two model architectures, analyzed the target speaker over-suppression (TSOS) problem, and proposed a metric to measure the TSOS. We also performed comprehensive evaluation by considering different scenarios and audio durations. Our results showed that the proposed models provided promising results for PSE in various conditions. Furthermore, we showed that the asymmetric loss function, multi-task training, and their combination would significantly mitigate the TSOS problem.

\vfill\pagebreak

\bibliographystyle{IEEEtran}
{\small\bibliography{refs}}

\end{document}